# Methodology for design of templates of text communication messages for software marketing

Elena K. Malakhovskaya[1], Yury P. Ekhlakov[1], Pavel V. Senchenko[1], Anatoly A.Sidorov[1]
[1]Tomsk State University of Control Systems and Radioelectronics
Tomsk, Russian Federation
elena_tusur@mail.ru

*Abstract* — **Communication messages play a key role in marketing of products and services, and the software market is no exception. A methodology is proposed for design of templates of text communication messages that are based on best practices of experts in software marketing, ideas of marketing, communication theory, copywriting, media linguistics, semiotics; description of the subject area is based on conceptual modeling and production systems. For the purposes of testing, the methodology was used as the basis of a software product – a decision support recommender system for design of communication messages for software marketing.**

## I. Introduction

The financial wellbeing of an IT company is in many ways determined by the quality of its promotional activities for marketing of its software products (SP) and / or services. It is crucial to develop effective marketing communications with potential consumers of the SP, their main goal being to not only make the SP visible to the target audience (TA) and convey its competitive advantages, but also to drive the TA to decide to try using it, and later purchase it.

The most difficult aspect of organizing marketing communications is creating the messages, i. e. the information that will be delivered to the potential consumer in order to direct their attention to the product (service) and prompt a certain response. When creating such messages, particular attention should be paid to the structure, content and presentation of the communication message (CM) as the main carrier of information about the functionality and usability of the SP.

Papers that discuss the methodological aspects of CM design through the lens of applied linguistics do so to identify and study the "rules" of developing effective marketing communications. The recent decades have seen the rise of media linguistics that studies the transformation of natural language texts for the purposes of the media sphere with the help of discourse analysis, functional stylistics, rhetoric, cognitive linguistics, etc. [1-4].

When discussing the appeal of the internet as a major channel for consumer communication, it is important to mention the publications by F. Yu. Virin, O. A. Kobelev, A. V. Yurasov, I. S. Ashmanov [5-8] that explore the aspects of internet marketing. It should be noted that the matters of marketing communications are most often discussed through cases and the personal experience of internet marketing experts shared on blogs and specialized forums, which practice, while having undeniable practical significance, does not allow for a systematic approach to marketing communications. It fails to pay any significant attention to the matters of software product promotion.

Based on the key branches of media linguistics, communication theory, semiotics, methods of artificial intelligence, ideas of object-oriented design and design patterns, this paper proposes a methodology for design of templates of text communication messages for software product promotion in the corporate market. The idea of a template as a tool for development and design was first described in [9], which discussed the concept of a template language as a "structured approach to describing effective design methods in the context of the current best practices".

## II. Methodology for design of text communication message templates

Let us represent the methodology for design of text CM templates for SP promotion as a sequence of five principal stages (Figure 1).

*Analyze the basic rules of CM design Abbreviations and Acronyms*

At the first stage, the subject area, i. e. communication messages, is described in the CM template design concept. Based on the rules for composition of advertising texts, including those described in [10-12], a model of a text communication message as a structured piece of information expressed in language units with a standard design (determined by the advertising platform) can be represented by a tuple of four interrelated parameters:

$$< C, F, V, S >,$$

Where C is the CM semantics (meanings of the communication message as a whole or its individual structural parts);

$F = \{F_i\}, i = \overline{1,n}$, is a set of stylistic devices that determine the format of the communication message (question, argument, invitation to action, etc.);

$V = \{V_j\}, j = \overline{1,m}$, is the number of symbols that communicate the CM;

S is the structure of the CM (tagline, title, main text, reference information, echo phrase) or another set that describes the structural parts that determine the logical sequence of language units in the CM).

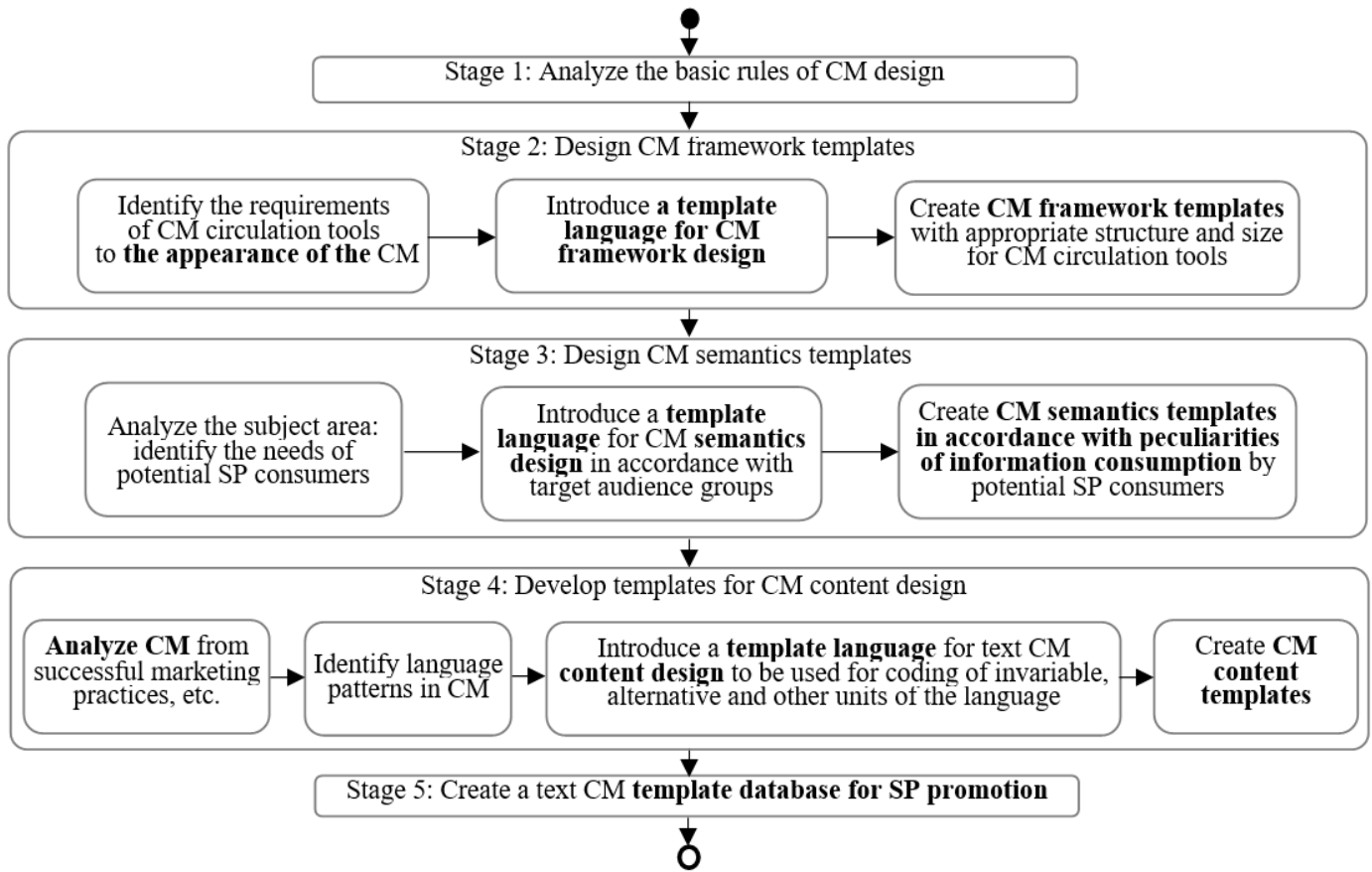

Fig. 1. Stages for design of text communication message templates

For example, it is important to understand that the title, as a structural element of the CM, has a certain meaning (function) and determines the logical sequence of language units – it draws the audience's attention to the SP. Its semantics (meaning) should focus on the main argument (unique selling point) that is disclosed in the main text or on address to the consumer.

At the same time, language units must have common points with the thesaurus of the recipient of information. The format can be set as a question, appeal to the problem, etc. The tool for CM circulation determines the size of the header, e.g. Google AdWords allows for 60 characters, while Yandex.Direct allows for 35 (with additional 30 characters).

Figure 2 shows factors that affect the values of CM parameters, which will be important to take into account at later stages of CM template design.

ACKNOWLEDGMENT

This paper is designed as part of the state assignment of the Ministry of Science and Higher Education; project FEWM-2020-0036.